\documentclass{lipics-v2021}

\usepackage[sort&compress,numbers]{natbib}

\let\epsilon=\varepsilon
\def\defn#1{\textbf{\textit{\boldmath #1}}}

\newcommand{\xxx}[1]{}

\title{Reconfiguration Algorithms for Cubic Modular Robots with Realistic Movement Constraints}

\titlerunning{Reconfiguration Algorithms for Cubic Modular Robots}

\author{MIT--NASA Space Robots Team\footnote{Artificial first author to highlight that the other authors (in alphabetical order) worked as an equal group. Please include all authors (including this one) in your bibliography, and refer to the authors as ``MIT--NASA Space Robots Team'' (without “et al.”).}}{Massachusetts Institute of Technology and NASA Ames Research Center}{}{}{}
\author{Josh Brunner}{Massachusetts Institute of Technology}{brunnerj@mit.edu}{}{}
\author{Kenneth C. Cheung}{NASA Ames Research Center}{kenny@nasa.gov}{}{}
\author{Erik D. Demaine}{Massachusetts Institute of Technology}{edemaine@mit.edu}{https://orcid.org/0000-0003-3803-5703}{}
\author{Jenny Diomidova}{Massachusetts Institute of Technology}{diomidova@mit.edu}{}{}
\author{Christine Gregg}{NASA Ames Research Center}{christine.e.gregg@nasa.gov}{}{}
\author{Della H. Hendrickson}{Massachusetts Institute of Technology}{della@mit.edu}{}{}
\author{Irina Kostitsyna}{KBR at NASA Ames Research Center}{irina.kostitsyna@nasa.gov}{https://orcid.org/0000-0003-0544-2257}{}

\authorrunning{MIT--NASA Space Robots Team}

\Copyright{Josh Brunner, Kenneth C. Chung, Erik D. Demaine, Jenny Diomidova, Christine Gregg, Della H. Hendrickson, and Irina Kostitsyna} 



\nolinenumbers 

\hideLIPIcs  

\begin{document}

\maketitle

\begin{abstract}
We introduce and analyze a model for
self-reconfigurable robots made up of unit-cube modules.
Compared to past models, our model aims to newly capture
two important practical aspects of real-world robots.
First, modules often do not occupy an exact unit cube, but rather have features like bumps
extending outside the allotted space so that modules can interlock.
Thus, for example, our model forbids modules from squeezing
in between two other modules that are one unit distance apart.
Second, our model captures the practical scenario of many passive modules
assembled by a single robot, instead of requiring all modules to be able
to move on their own.

We prove two universality results.
First, with a supply of auxiliary modules, we show that any connected polycube structure can be constructed by a carefully aligned plane sweep.
Second, without additional modules, we show how to construct any structure for which a natural notion of external feature size is at least a constant; this property largely consolidates forbidden-pattern properties used in previous works on reconfigurable modular robots.
\end{abstract}

\section{Introduction}



Algorithmic shape formation with self-reconfigurable modular robots has attracted significant interest by the computational geometry community in the past two decades~\cite{Dumitrescu2006,pushing-cubes,Crystalline_WAFR2008,Crystalline_ISAAC2008,Crystalline_CGTA,squeezing11,MTRAN_Molecube_CGTA,M-blocks,Feshbach2021,Lo-Wong2021,Parada2021,pivoting-socg21,musketeers,Akitaya2022,Chiang01,heuristics-square,Benbernou2011,PSPACE-sliding-corners,heterogeneous-03,flooding,BachelorMoreno}.
In general, the idea is to build a self-reconfigurable ``robot'' out of $n$
identical modules, each of which can move in some way relative to its neighbors,
subject to some constraints like maintaining global connectivity.
This approach enables the robot to drastically change its overall shape,
often with algorithmic universality results showing that any shape is possible,
up to some constant feature size and/or forbidden small patterns.
Real-world modular robots have been built by multiple robotics groups
\cite{M-blocks-practical,EMCube,Telecube,Vertical98,Crystal,survey2007},
with the ultimate goal of building ``programmable matter'':
objects that can arbitrarily change their shape.

\begin{figure}[h]
\begin{minipage}[t]{0.47\textwidth}
	\centering
	\includegraphics[scale=0.75]{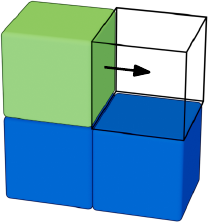}
	\qquad
	\includegraphics[scale=0.75]{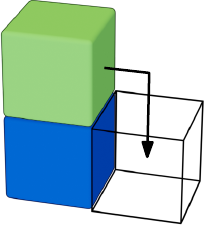}
	\caption{The two moves in the sliding-cubes model.
    Left: straight slide. Right: corner slide.}
	\label{fig:moves}
\end{minipage}
\hfill
\begin{minipage}[t]{0.49\textwidth}
	\centering
	\includegraphics[scale=0.75]{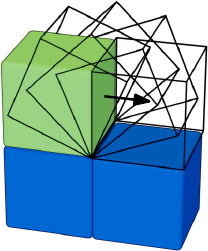}
	\qquad
	\includegraphics[scale=0.75]{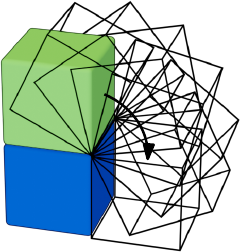}
	\caption{The two moves in the pivoting-cubes model.
    Left: straight pivot. Right: corner pivot.}
	\label{fig:pivot-moves}
\end{minipage}
\end{figure}

\subparagraph{Sliding cubes.}
One of the first and simplest models for modular robot reconfiguration is
\defn{sliding squares} in 2D \cite{Dumitrescu2006} or \defn{sliding cubes}
in 3D \cite{pushing-cubes,sliding-cubes-quadratic}.
Each module is a unit square/cube placed at a node of a square/cube lattice,
and the modules must at each step have a connected dual graph
(according to facet adjacencies).
Figure~\ref{fig:moves} illustrates the two possible moves:
(1)~\defn{straight slide} moves a (green) module to an adjacent empty location along two faces of a pair of adjacent (blue) modules; and
(2)~\defn{corner slide} moves a (green) module to an adjacent empty location, and immediately turning the $90^\circ$ corner around its original (blue) neighbor, moves the module one more unit to restore the adjacency with the neighbor.
This model enables universal reconfiguration between polyominoes/polycubes,
with $\Theta(n^2)$ moves necessary and sufficient in the worst case
for 2D \cite{Dumitrescu2006}
and (in a recent breakthrough) for 3D \cite{sliding-cubes-quadratic, sliding-cubes-in-place-quadratic}.

\subparagraph{Pivoting cubes.}
Another extensively studied model for modular robot reconfiguration is the \defn{pivoting squares/hexagons} in 2D~\cite{Benbernou2011,musketeers,pivoting-socg21} and \defn{pivoting cubes} in 3D~\cite{M-blocks,Feshbach2021,Lo-Wong2021}.
In the pivoting-cube model, a module can move by rotating around a common edge shared with an adjacent module.
Similarly to the previous model, a module moves to an adjacent empty location or to an empty location around the corner of an adjacent module (see Figure~\ref{fig:pivot-moves}).
However, unlike in the sliding-cube model, for a pivoting move to be valid, all cells of the grid intersected by the pivoting module must remain empty.
Existing results rely on the definition of so called \defn{forbidden pattern}, that is, a specific constant-size configuration of empty and non-empty cells, whose existence may block a possible reconfiguration.
Specifically, the forbidden patterns are of the form: for any $k_1\times k_2 \times k_3$ (for specific small values of $k_1$, $k_2$, and $k_3$) axis aligned bounding box of grid cells with two modules present in the two diagonally opposite corners of the box, the remaining cells of the box must not be empty.
The set of forbidden patterns consists of the $3\times1\times1$-pattern (two modules with a single empty cell in between), the $2\times2\times1$-pattern (two edge adjacent modules with no other mutually adjacent modules), and the  $3\times2\times1$-pattern.
The series of works on reconfiguration in the pivoting-cube model~\cite{M-blocks,Feshbach2021,Lo-Wong2021} resulted in a universal reconfiguration algorithm for a class of shapes that do not contain any of the three forbidden patterns.

\subsection{Our Model}
In this paper, we introduce a new model for modular robot reconfiguration
that aims to capture two important practical aspects of real-world systems,
motivated by our experience with the robots and structural modules of the NASA Ames Automated Reconfigurable Mission Adaptive Digital Assembly
Systems (ARMADAS) project \cite{park_solle_2023}.
Our model is a refinement of the sliding-cubes model,
adding constraints to the moves.
Notably, the constraints on the moves in our model are strictly stronger than in the pivoting-cube model as well.
Thus our universality results can be seen as strengthenings of past work
to better apply to real-world robotics.

\begin{figure}[t]
	\centering
	\includegraphics[width=1.0\textwidth]{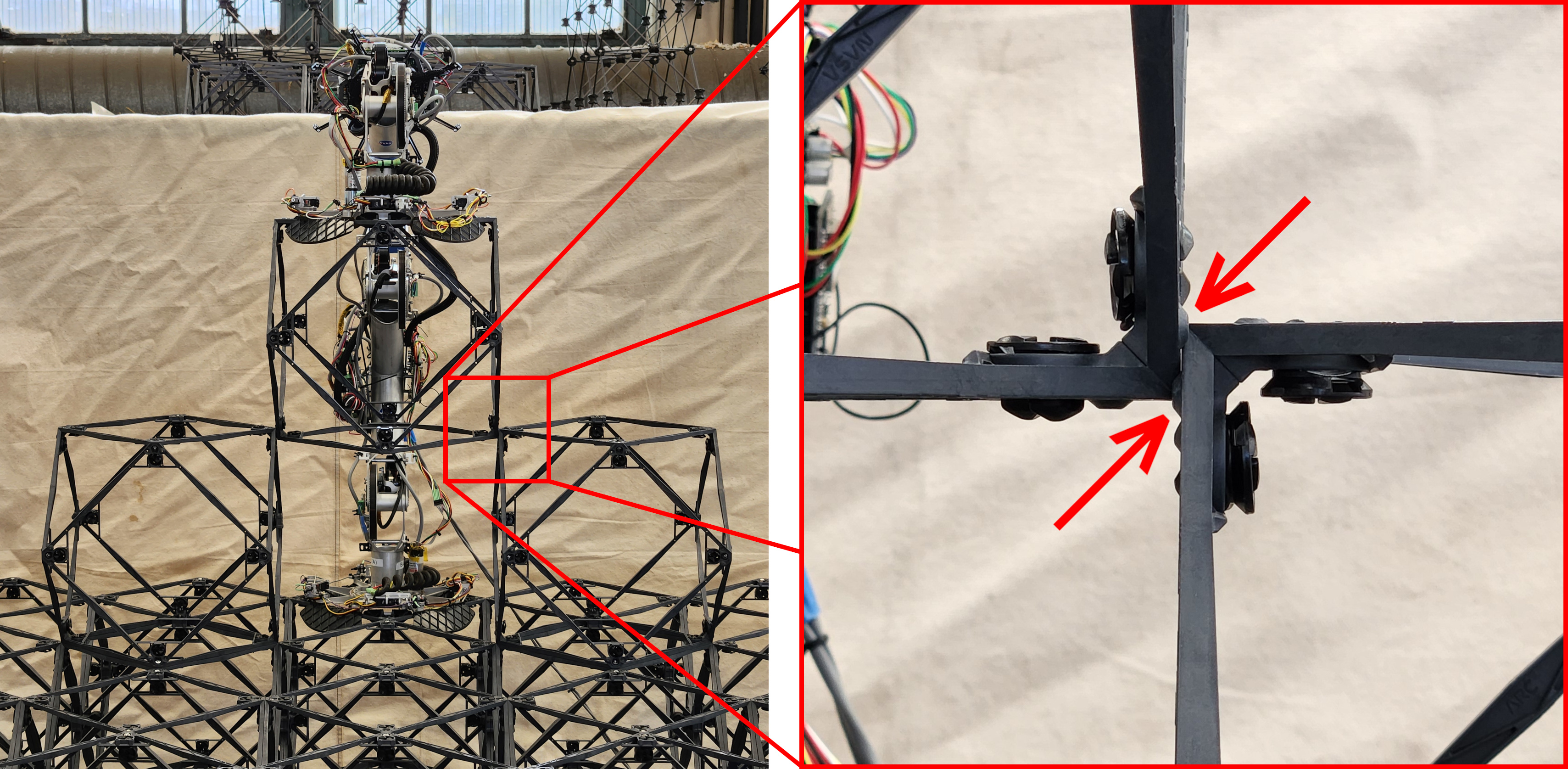}
	\caption{Photograph of ARMADAS robot attempting to place a module between two modules and failing due to collisions of mechanical alignment features (red arrows).}
	\label{fig:collisionPictures}
\end{figure}

\subparagraph{Loose sliding.}
The first practical issue we address is that modules typically
need more room than a unit cube to actually move without collision.
Figure~\ref{fig:collisionPictures} shows an example of this issue
in the context of ARMADAS.
To enable secure and precise relative positioning of adjacent modules,
modules often have mechanical alignment features
(matching bumps and indentations) that extend outside the bounding box.
Having these features integrated into modules allows for high-precision,
high-repeatability, and high-throughput manufacturing processes such as
injection molding to make high-quality connections,
rather than requiring the added complexity and weight of active attachment
mechanisms \cite{Cheung_reversible_2013, Gregg_ultralight_2018}. 

\begin{figure}[t]
	\centering
	\includegraphics[scale=0.75]{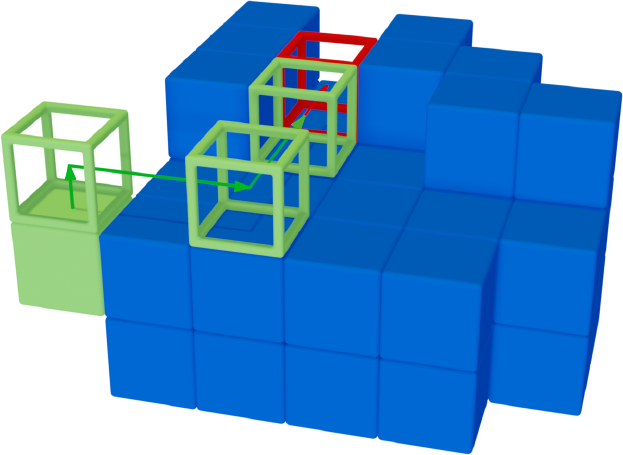}
	\caption{Valid (green) and invalid (red) moves of the green module
    in the loose-sliding model.}
	\label{fig:motivation}
\end{figure}

To avoid collision between modules with these bumps,
modules need to move slightly away
from neighboring modules before sliding to an adjacent cell.
Thus a moving module needs the space to move a small distance away from its neighbors.
In particular, it is impossible for a module to pass through a unit-wide
gap between two other modules; see Figure~\ref{fig:motivation}.

\begin{figure}[t]
	\centering
	\includegraphics[scale=0.75]{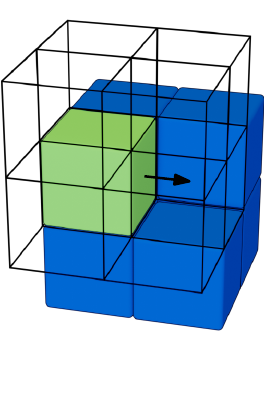}
	\hfill
	\includegraphics[scale=0.75]{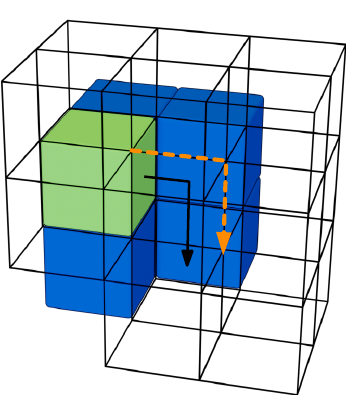}
	\hfill
	\includegraphics[scale=0.75]{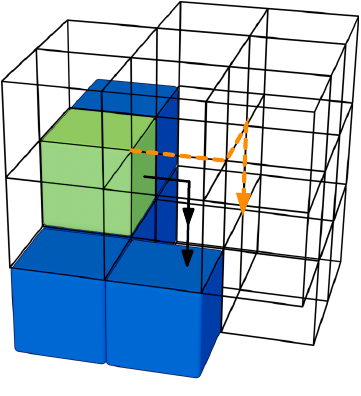}
	\caption{Valid moves in the loose-sliding model. Left: straight slide move and the corresponding $2\times 2\times 2$ empty cube. Middle and right: corner slide move and the corresponding translation (indicated by the orange paths) of the $2\times 2\times 2$ empty cube.}
	\label{fig:moves2}
\end{figure}

We formalize this requirement as the \defn{loose-sliding} constraint
(refer to Figure~\ref{fig:moves2}):
a moving module must at all times be within an otherwise empty
$2 \times 2 \times 2$ cube in space that moves continuously
(or equivalently, moves in unit axis-aligned steps on a grid).
In other words, each unit step taken by the moving module
(one for a straight slide, two for a corner slide)
must have both its start and end positions
within a common $2 \times 2 \times 2$ cube empty of other modules;
in addition, in the case of a corner slide, the first $2 \times 2 \times 2$ empty cube must be translatable
(continuously or in unit axis-aligned steps)
to the second $2 \times 2 \times 2$ empty cube while remaining empty of other modules throughout the translation.
In particular, loose sliding prevents a module from sliding into a unit-wide
gap between two other modules, because such a gap is not contained in an empty
$2 \times 2 \times 2$ cube.
More generally, we can define \defn{$k$-loose  sliding} to require a
$k \times k \times k$ empty cube surrounding the moving module.

\subparagraph{Passive modules via accessible sliding.}
The second practical issue is that modular re\-con\-fig\-urable robots do not scale well.
Each module must have its own power, processor, networking,
attachment mechanisms, and movement actuators.
This module complexity limits the number of modules and practicality of the
programmable matter dream.

\begin{figure}[t]
	\centering
	\includegraphics[width=0.7\textwidth]{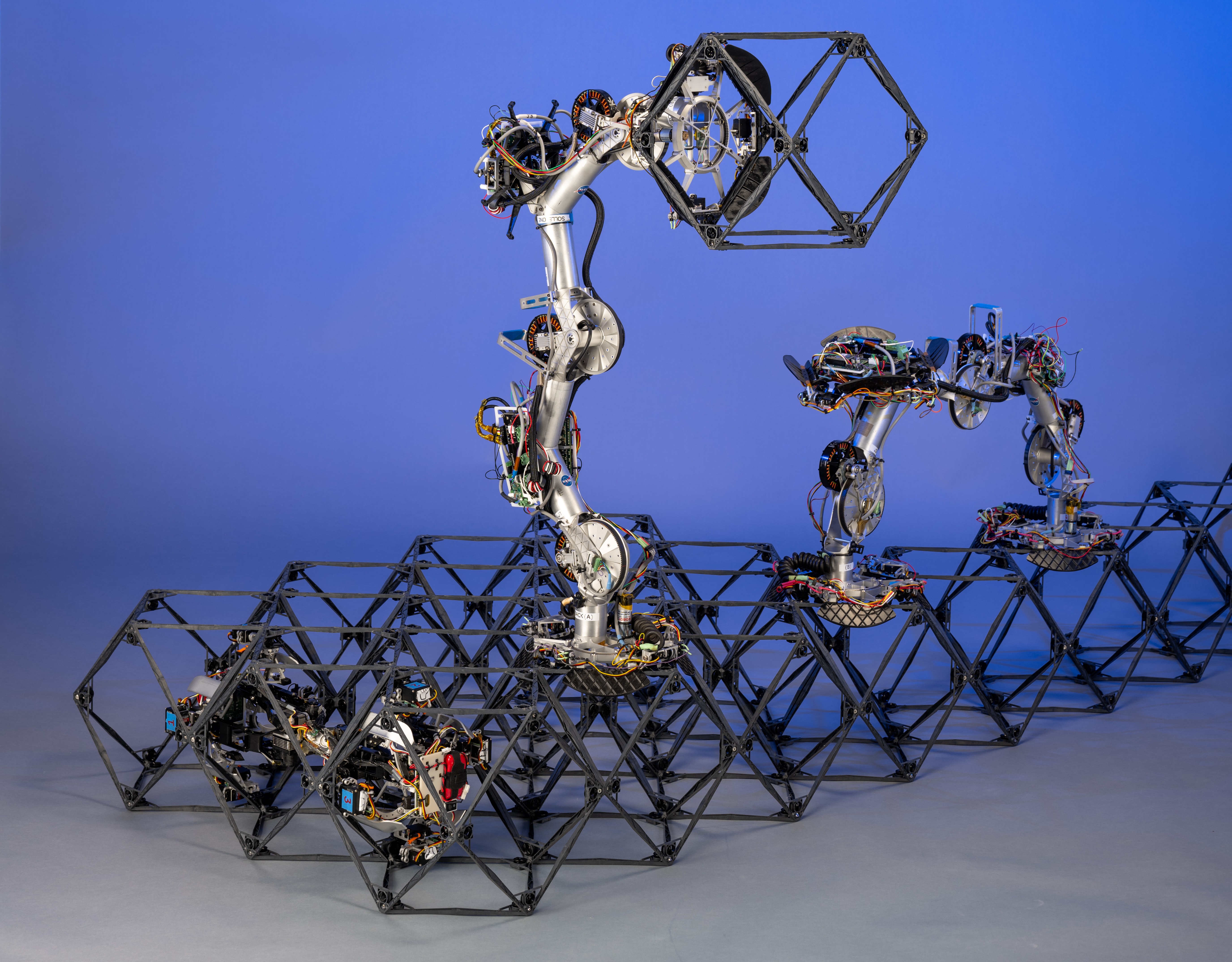}
	\caption{Photograph of ARMADAS robots operating in the laboratory.
    Each of the external robots has two primary ``hands'', and can hold onto the already-built
    structure with either hand while moving the other hand.
    The right robot is carrying a module via a third hand on its back;
    this module can be added to the structure by another robot,
    such as the left robot.}
	\label{fig:robotPicture}
\end{figure}

A more recent alternative approach \cite{park_solle_2023,jenett_materialrobot_2019,terada_automatic_2005} is to have two types of modules:
many passive/static modules that cannot move on their own and
primarily serve structural purposes, and
a smaller number of robots (even one) that can pick up, carry, place,
and attach passive modules.
Figure~\ref{fig:robotPicture} shows this approach in action
in the context of the ARMADAS project, where the primary goal is to assemble
passive parts into a desired geometry --- a digital cellular solid \cite{Cheung_reversible_2013, armadas}.
(Of course, disassembly and reconfiguration is also possible.)
The moving robots are more complicated than the passive modules,
and thus are naturally a little larger.
The simplicity of passive modules enables them to be cheaply constructed
in large numbers, vastly increasing applicability \cite{Gregg_ultralight_2018}.

Instead of explicitly modeling both moving robot(s) and passive modules,
which would vastly complicate the model,
we show how to obtain a similar effect via a small tweak to the
loose-sliding-cubes model.
Specifically, define a slide to be \defn{$k$-accessible} if it is both $k$-loose (i.e. is the only module in some otherwise empty $k \times k \times k$ cube), and also that that $k \times k \times k$ is connected to infinity via a path of $k \times k \times k$ empty cubes,
both before and after the move.
Such a module can be reached by a moving robot on the outside of the shape,
removed and picked up by the robot, and then placed and attached
in the new location.
Thus any sequence of $k$-accessible moves can be simulated by passive modules
plus one moving robot whose size is at most $k \times k \times k$.

Our algorithms for sliding cubes satisfy this $k$-accessible property.
Thus they are equally suitable for both modular robot reconfiguration
(where every module can move on its own)
and a hybrid system of passive modules and one or more moving robots.
By contrast, all past sliding-cube algorithms
\cite{Dumitrescu2006,pushing-cubes,sliding-cubes-quadratic}
do not satisfy the accessible property.
Indeed, \emph{no} universal algorithm without extra modules can be accessible
(even $1$-accessible), because it is known that there are 3D configurations
with no movable modules on the outside \cite{cubes-multimedia}.

\subsection{Our Results}

We develop two very different reconfiguration algorithms,
establishing two different universality results
in the $2$-accessible sliding-cubes model:
\begin{enumerate}
\item
  Allowing extra modules, we show how to construct \emph{any} connected
  polycube from a straight line, and thus how to reconfigure between
  any two connected polycubes [Section~\ref{sec:extra}].
  The total number of extra modules is linear in the number of modules in the polycube. We also give a negative result that there are structures which are impossible to reconfigure between in the $3$-loose sliding-cubes model, thus showing that our algorithm's result is tight.
\item
  Without extra modules, we show how to construct any connected polycube
  having ``external feature size'' at least~$2$,
  and thus how to reconfigure between any two such polycubes
  [Section~\ref{sec:no-extra}].
%
Here we define \defn{external feature size at least $k$} as follows:
\begin{enumerate}
\item Every empty $1 \times 1 \times 1$ cube $Q$
  is contained in an empty $k \times k \times k$ cube $\hat Q$ (see Figure~\ref{fig:moves2} (left)); and
\item For every pair of edge-adjacent empty $1 \times 1 \times 1$ cubes
  $Q_1$ and $Q_2$, there exist empty $k \times k \times k$ cubes
  $\hat Q_1$ and $\hat Q_2$ containing $Q_1$ and $Q_2$ respectively,
  such that we can continuously slide $\hat Q_1$ into $\hat Q_2$
  by axis-parallel unit slides while preserving emptyness of the
  intermediate $k \times k \times k$ cube and remaining within the
  bounding box of $\hat Q_1 \cup \hat Q_2$ (see Figure~\ref{fig:moves2} (middle and right)).
\end{enumerate}
\end{enumerate}

As already mentioned, $2$-loose sliding is a slightly stricter constraint than in the pivoting-cubes model.
In our second result we build upon the techniques developed for both, the sliding-cube and the pivoting-cube models.

The second construction has a useful additional property called
\defn{monotonicity}.
Assume we start from a line or box of modules, and the goal is to
assemble the desired polycube adjacent to this starting configuration.
Then we move the modules in order, and once a module stops moving,
we never move it again.
Monotonicity implies both accessibility and the lack of extra modules.
Furthermore, monotonicity tells us that the assembly process is particularly
efficient to simulate with one moving robot and many passive modules:
the robot simply needs to double the motion of each module
(once to move the module in place, and then in reverse to get the next module).

Our algorithm descriptions focus on the case $k=2$,
as it is the simplest case where the model differs from past work,
and as an ARMADAS robot can fit within a $2 \times 2 \times 2$ cube.

\section{Universality with Extra Modules: 3D Printing}
\label{sec:extra}


In this section, we will give a 2-accessible algorithm to disassemble any connected structure using extra ``scaffolding'' modules. More precisely, given an initial configuration $T$ of $n$ modules, and a line of $O(n)$ extra modules attached to it, the algorithm will reconfigure the structure into a single line of modules. The intuition behind the algorithm is to ``3D print'' the structure in reverse by gradually moving a sweep plane through the structure. The sweep plane is filled with modules to preserve the connectivity of the structure, allowing the modules of $T$ to be removed one at a time.
We will first describe a simpler version of the algorithm for two-dimensional structures to give intuition.

\subsection{2D}


Let $T \subseteq \mathbb{Z}^2$ be the shape we want to disassemble and $U \subseteq \mathbb{Z}^2$ be an axis-aligned rectangle which contains $T$. We will reconfigure it into a single long line $L$.

The main idea is to sweep $U$ with a diagonal line of scaffolding, placing modules in front of it if they are not already present, and removing modules behind it. We use a possibly counterintuitive slope of $3:5$ for the sweep line.
As it will become clear later, it is hard to maintain connectivity of the intermediate structure with simpler slopes. In particular, a simpler horizontal, vertical or $1:1$ diagonal slope does not allow us to ``dig'' a small hole in the sweep line to remove a module from behind it without either temporarily breaking connectivity or moving a module that is not 2-accessible.

\begin{figure}
    \centering
    \includegraphics[scale=0.3]{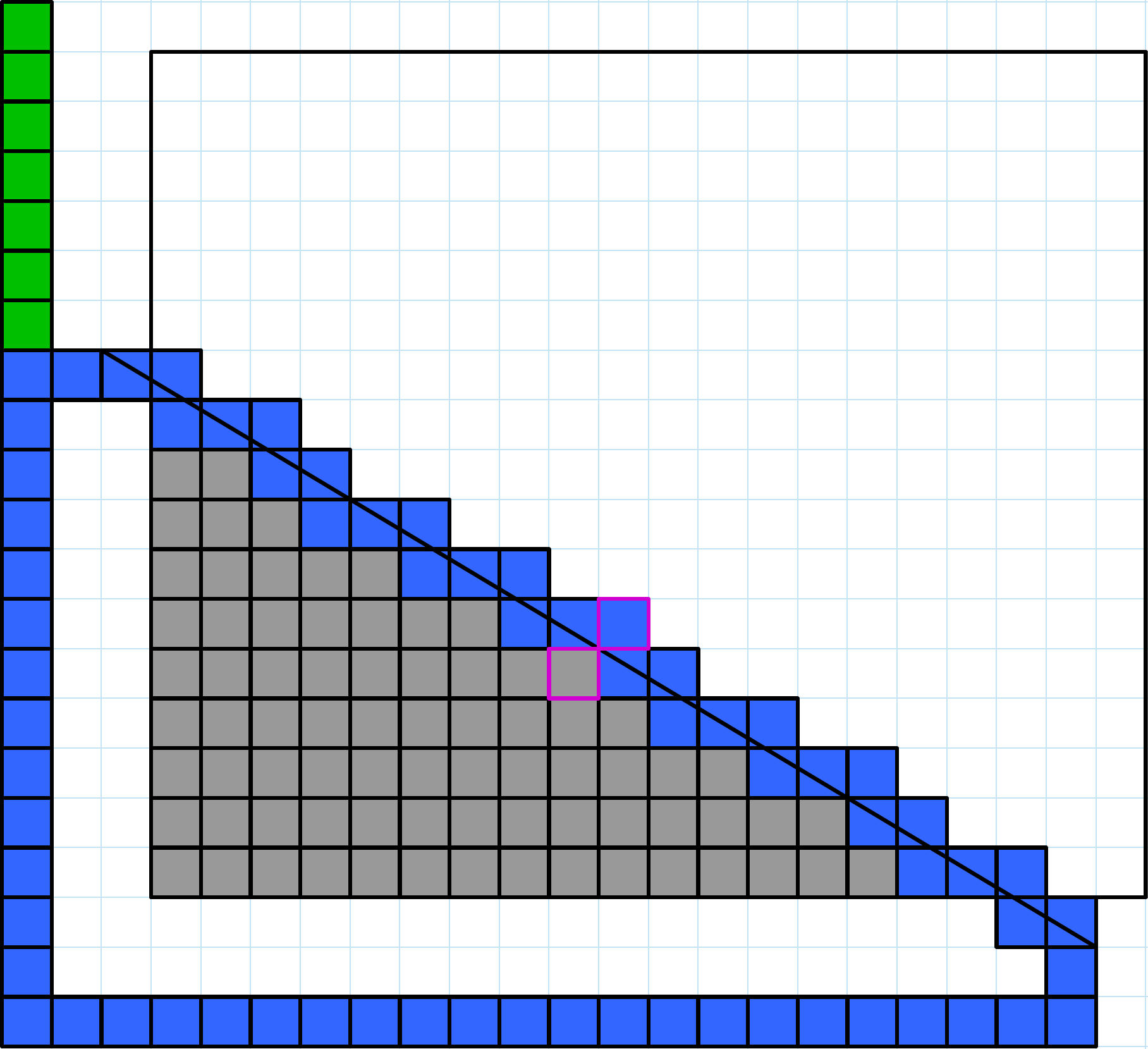}
    \caption{The 2D sweep line approach. Blue squares are part of the sweep line, gray squares represent the remainder of $T$ that has not been deconstructed. (Not all gray squares correspond to existing modules, since $T$ is an arbitrary connected subset.) Purple squares are the next modules to be added or removed. The outer black box is the boundary of $U$. The green squares are the line $L$ where removed modules are appended.}
    \label{fig:2d-S_t}
\end{figure}

Let $f(x,y) = 3x + (5 - \epsilon)y$ for some small $\epsilon \ll 1/n$. Our diagonal line of scaffolding will consist of all modules which are intersected by the line $f(x,y) = t$ for some $t$ which we gradually decrease to sweep the line across. This is equivalent to sweeping a $3:5$ slope line across, and when it hits several modules simultaneously, tiebreaking in favor of processing the higher $y$ valued ones first. We define an intermediate state $R_t$ to be the union of the following four sets (refer to Figure~\ref{fig:2d-S_t}):

\begin{itemize}
    \item remainder of the structure $\{(x,y) \in T \mid f(x,y) \le t\}$,
    \item a diagonal ``sweep line'' of modules $\{(x,y) \in U \mid t < f(x,y) \le t + 8 - \epsilon\}$,
    \item a path outside of $U$ connecting the two ends of the sweep line. This is just to ensure the structure remains connected when we remove modules from the sweep line, and
    \item the portion of $L$ that has been built so far: a line of modules that have been removed from $T$ that is attached to the path outside of $U$.
\end{itemize}

\begin{observation}
    Any 2-accessible module can slide along the exterior of $R_t$ to the end of the line $L$ in $O(n)$ moves.
\end{observation}

Using the above observation, any extra modules that are not part of $R_t$ at time $t$ are placed in the line $L$. For the rest of this section, whenever we refer to ``removing'' a module, we mean sliding it to the end of $L$.

Initially, we construct a bounding box made of the additional modules around $U$ with an extra gap of two units.
Furthermore, we connect $T$ to this bounding box.
Specifically, we build a path of modules out from the right-most top-most module of $T$ to the upper right corner of the bounding box. This is done simply to ensure the sweep line is connected to $T$ initially (and we can equivalently think of $T$ as containing this additional path).

We are going to gradually decrease $t$, starting from $R_{\infty} = T$ and ending at $R_{-\infty} = \emptyset$. As we decrease $t$, $R_t$ changes only when $t=f(x,y)$ or $t=f(x,y)-8+\epsilon$ for some $(x,y) \in U$, which happens $|U|$ times. More specifically, a module at $(x,y)$ is added to the sweep line when $t=f(x,y)$, and is removed from it when $t=f(x,y)-8+\epsilon$. Note that, whenever the first condition is met for $(x,y)$, the second condition is met for $(x+1,y+1)$. Thus, each event updating $R_t$ will involve removing one module at $(x+1,y+1)$ and adding one module at $(x,y)$. Consecutive states of $R_t$ differ in only these two modules. We will show that we can always transition from $R_t$ to $R_{t-\epsilon}$ in a constant number of moves near the sweep line, plus, if necessary, a linear number of moves to take the removed modules out to the end of the line $L$.

\begin{lemma}
There is a sequence of $O(n)$ 2-accessible moves that reconfigures $R_t$ to $R_{t-\epsilon}$.
\end{lemma}
\begin{proof}
Consider a sweep-line event $t=f(x,y)$ for some $(x,y)$. When processing this event, $R_t$ changes in two ways: a module is removed from $(x+1,y+1)$ and a module is added (if needed) at $(x,y)$. If $(x,y)$ is outside the boundary of $U$, all the changes that need to happen are sufficiently far from $T$, so we can simply deconstruct the sweep line one module at a time starting from the boundary of $U$ and then rebuild it in its new configuration. The structure will always stay 2-accessible since $T$ is at least two units distance away from the moving modules and thus cannot break the $2$-loose sliding property.

The interesting case is when $(x,y)$ is inside the boundary of $U$, as shown in Figure~\ref{fig:2d-S_t}, where the two cells outlined in purple are being processed: we need to place a module at $(x,y)$ (if it is absent) and remove a module at $(x+1,y+1)$. Removing the latter is trivial since it is on the surface of $R_t$. It suffices to show how to place a module at $(x,y)$ by ``digging through'' the line of scaffolding.
The idea is to carefully make a small hole in the scaffolding, place a module at $(x,y)$ through it, and then seal the hole. The challenge is to do this while preserving connectivity.

\begin{figure}
    \centering
    \includegraphics[scale=0.3]{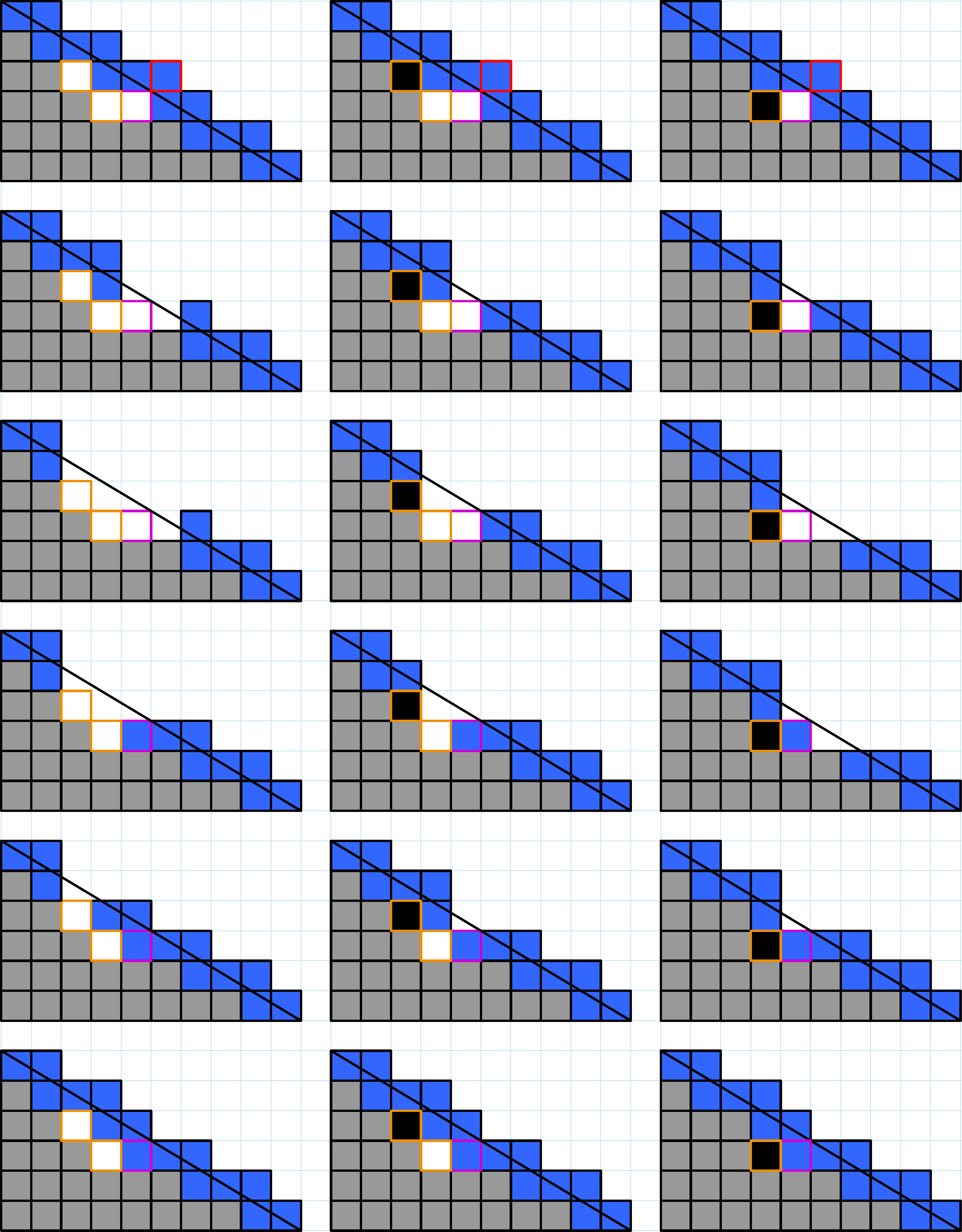}
    \caption{Three cases of processing a sweep-line event. Reading from top to bottom, each column shows the intermediate steps of adding and removing a module from $R_t$ in the corresponding case. The module at $(x+1,y+1)$ (outlined in red) is being removed and a module is being added at $(x,y)$ (outlined in purple). Case 1 (left column): two cells $(x-1,y)$ and $(x-2,y+1)$ (outlined in orange) are empty. Case 2 (middle column): cell $(x-1,y)$ is empty but cell $(x-2,y+1)$ is occupied. Case 3 (right column): cell $(x-1,y)$ is occupied.}
    \label{fig:2d-cases}
\end{figure}

The exact sequence of steps is shown in Figure~\ref{fig:2d-cases}, with three cases depending on whether there are modules at $(x-1,y)$ and $(x-2,y+1)$ (shown with orange outline). 

Case 1: There is a module at $(x-1,y)$  (top row of Figure~\ref{fig:2d-cases}):
\begin{enumerate}
    \item Remove modules at $(x+1,y+1)$ and $(x,y+1)$.
    \item Remove modules at $(x+1,y)$ and $(x+2,y)$.
    \item Place a module at $(x,y)$.
    \item Place modules at $(x+1,y)$ and $(x+2,y)$.
    \item Place modules at $(x,y+1)$.
\end{enumerate}

Case 2: There is a module at $(x-2, y-1)$ but no module at $(x-1,y)$ (middle row of Figure~\ref{fig:2d-cases}):
\begin{enumerate}
    \item Remove modules at $(x+1,y+1)$ and $(x,y+1)$.
    \item Remove modules at $(x-1,y+2)$ and $(x-1,y+1)$.
    \item Place a module at $(x,y)$.
    \item Place modules at $(x-1,y+1)$ and $(x-1,y+2)$.
    \item Place modules at $(x,y+1)$.
\end{enumerate}

Case 3: There is no module at $(x-1,y)$ and $(x-2, y-1)$ (bottom row of Figure~\ref{fig:2d-cases}):
\begin{enumerate}
    \item Remove modules at $(x+1,y+1)$ and $(x,y+1)$, $(x+1,y)$.
    \item Remove modules at $(x-1,y+1)$, $(x-1,y+2)$, and $(x-2,y+2)$.
    \item Place modules at $(x+1,y)$ and $(x,y)$.
    \item Place modules at $(x,y+1)$ and $(x-1,y+1)$.
    \item Place modules at $(x-2,y+2)$ and $(x-1,y+2)$.
\end{enumerate}

The key point is that each of the individual placements and removals in each of these sequences is always accessible from the outside: each removed or placed module is the only occupied space of some $2\times 2$ square which is reachable from the outside. Thus, the algorithm is 2-accessible.

Furthermore, with each placement or removal of a module the connectivity of the structure is preserved. Whenever we removed a module from or scaffolding, any adjacent modules of the original structure still always had an adjacent to a module of scaffolding, thus ensuring that all of the remaining modules of the original structure remain connected to the scaffolding. Additionally, our scaffolding layer is internally connected throughout the operation, so the entirety of $S_t$ is connected.
\end{proof}

Because each dimension of $U$ is $O(n)$, the sweep line consists of $O(n)$ modules, so we only use $O(n)$ additional modules to deconstruct $T$.

\begin{theorem}
    Any polyomino shape $T$ can be deconstructed into a line with $2$-accessible sliding moves with the help of additional $O(n)$ modules.
\end{theorem}

\subsection{3D}

Next we describe the full three-dimensional algorithm for disassembling a structure. Let $T \subseteq \mathbb{Z}^3$ be the shape we want to disassemble, and $U \subseteq \mathbb{Z}^3$ be an axis-aligned bounding box which contains $T$.
Similarly to the 2D case, we sweep a diagonal plane of scaffolding. At each step, we add a module in front of the sweep plane if it is not already present and remove a module from behind the plane.

Define $f(x,y,z) = x + (3-\epsilon) y + (3-\epsilon^2) z$, for sufficiently small $\epsilon$. Again, we define $R_t$ to be the set of modules still present in the structure at time $t$, which is the union of three sets:
\begin{itemize}
    \item the remainder of the structure: $\{(x,y,z) \in T \mid f(x,y,z) \le t\}$,
    \item the diagonal sweep plane of scaffolding keeping everything connected: $\{(x,y,z) \mid t < f(x,y,z) \leq t+4-\epsilon\}$,
    \item the portion of $L$ that has been built so far: a line of modules that have been removed from $T$ that is attached to the path outside of $U$, and
    \item additional modules near the boundary of the sweep plane which help preserve connectivity of the sweep plane near the boundary of $U$: $\{(x,y,z) \mid t < f(x,y,z) \leq t+6-2\epsilon\}$, when $(x,y,z)$ is outside of $U$ but is within four-unit distance of the boundary of $U$.
\end{itemize}

Similarly to the 2D case, as the sweep plane moves, we keep the line $L$ attached to the scaffolding. Whenever a module is removed, it can slide along this connection to the end of $L$ to be deposited.

\begin{lemma}
There is a sequence of $O(n)$ 2-accessible moves that reconfigures $R_t$ to $R_{t-\epsilon}$.
\end{lemma}
\begin{proof}
As before, we continuously decrease $t$. Each time $R_t$ changes, up to one module is added to the sweep plane at $(x,y,z)$ and the module at $(x+1,y+1,z)$ is removed from it. Removing modules is straightforward: they can slide to the end of $L$. Adding modules, however, requires some care. When we need to add a module at location $(x,y,z)$ in front of the sweep plane, we `dig' a small hole in the plane, add the new module, and then fill in the hole back.

\begin{figure}[t]
\centering	
\includegraphics{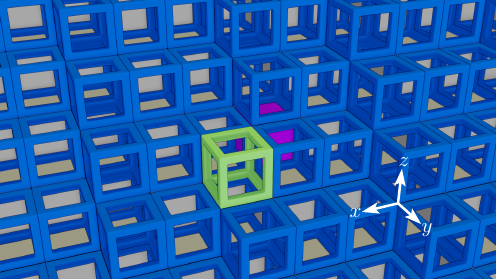}
\caption{The state of our ``3D printing'' algorithm before processing an event. The scaffolding plane is shown in blue. Purple cell is at $(x,y,z)$ and a module is to be placed in it; green module is at $(x+1,y+1,z)$ and is to be removed.}
	\label{fig:3dprint}
\end{figure}

\begin{figure}
\begin{minipage}[t]{0.48\textwidth}
    \centering
\includegraphics[width=\textwidth]{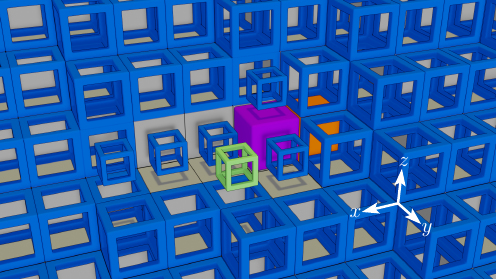}
    \caption{The solid orange module behind the purple cell is part of $T$. The small blue and green blocks correspond to the modules that are removed in order to access the purple cell without breaking connectivity.}
    \label{fig:3dprint-case1}
\end{minipage}
\hfill
\begin{minipage}[t]{0.48\textwidth}
    \centering
\includegraphics[width=\textwidth]{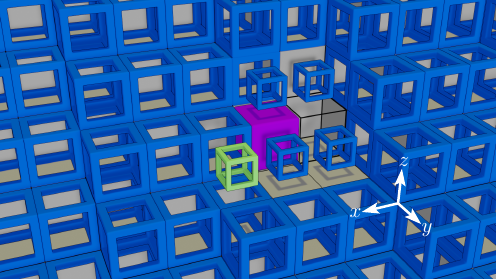}
    \caption{There is an empty cell behind the purple cell. The small blue and green blocks correspond to the modules that are removed in order to access the purple cell without breaking connectivity.}
    \label{fig:3dprint-case2}
\end{minipage}
\end{figure}

Next we provide specific steps for processing the event $f(x,y,z)=t$.
First, if a module at $(x,y,z)$ already exists in $R_t$, we simply remove the module at $(x+1,y+1,z)$.
When a cell $(x,y,z)$ is not occupied and a module needs to be placed in it, we consider the following two cases, based on whether $(x-1,y,z)$ is present or not.
\xxx{Journal version: Try to add figures/animation of 3D intermediate steps.}

Case 1: cell $(x-1,y,z)$ is occupied. See~Figure~\ref{fig:3dprint-case1}.
\begin{enumerate}
    \item Remove $(x+i,y+1,z)$ for $i = 1,0$,  and $(x,y,z+1)$.
    \item Remove $(x+i,y,z)$ for $i = 1,2,3$.
    \item Place $(x+i,y,z)$ for $i = 0,1,2,3$.
    \item Place $(x,y+1,z)$ and $(x,y,z+1)$.
\end{enumerate}

Case 2: cell $(x-1,y,z)$ is empty. See Figure~\ref{fig:3dprint-case2}.
\begin{enumerate}
    \item Remove $(x+i,y+1,z)$ for $i = 1,0,-1$, and $(x+i,y,z+1)$ for $i = 0,-1$.
    \item Place $(x,y,z)$.
    \item Place $(x+i,y+1,z)$ and $(x+i,y,z+1)$ for $i=-1,0$.
\end{enumerate}

Note, that every module that is being removed or added in the steps above is 2-accessible by construction.

When we are outside of $U$, we will use a different method to add and remove modules from the sweep plane. As the modules of the scaffold that are outside of $U$ are sufficiently removed from $T$, the empty cells in front of the sweep plane adjacent to them are in fact $2$-accessible. Thus, we can simply add a module directly to the front of the sweep plane and remove a module directly from the back side.

Observe that in this case the connectivity of the scaffolding is preserved. Every point $(x,y,z)$ with $f(x,y,z)<t$ that is face-adjacent to a module in the sweep plane maintains the property that at least one of the modules in the sweep plane that it is face-adjacent to is present at all points during the process. This means that if the structure was connected before this step of the algorithm, it remains connected during this step, and will still be connected afterward.
\end{proof}

Each dimension of $U$ is $O(n)$. However, because the sweep plane is two-dimensional, this means that our algorithm requires $O(n^2)$ additional modules. We can reduce the number of modules needed by reducing the size of $U$. Instead of using a full bounding box, we construct $U$ as follows. Consider a discretization of the lattice grid into $2\times 2\times 2$ metacells. For each module in $T$, consider the line passing through the center of the module perpendicular to the sweep plane. We let $U$ be the intersection of the bounding box of $T$ and union of all metacells that are pierced by any such line.

Because we maintain the $4$-module offset of the boundary of $U$ padded with an extra layer of modules on the front of the sweep plane, our sweep plane stays connected at each step of the reconfiguration regardless of the shape of the boundary of $U$. By construction of $U$, as it is made of $2\times2\times2$ metacells, the empty cells, adjacent to the front of the sweep plane, within distance $4$ of the boundary of $U$ are necessarily $2$-accessible. Since the original algorithm does not depend on the specific shape of $U$ but only the fact that the edge modules are $2$-accessible, the correctness of the algorithm follows.
The size of the intersection of our sweep plane with $U$ is $O(n)$ by construction, so we now only need $O(n)$ additional modules.

\begin{theorem}
    Any polycube shape $T$ can be deconstructed into a line with $2$-accessible sliding moves with the help of additional $O(n)$ modules.
\end{theorem}

\subsection{Impossibility of 3-Loose Algorithm}

In this section we describe a structure which cannot be reconfigured into a line with any 3-loose algorithm. We will first describe a 2D version and then generalize to a 3D version.

\begin{figure}[b]
	\centering
	\includegraphics[scale=0.3]{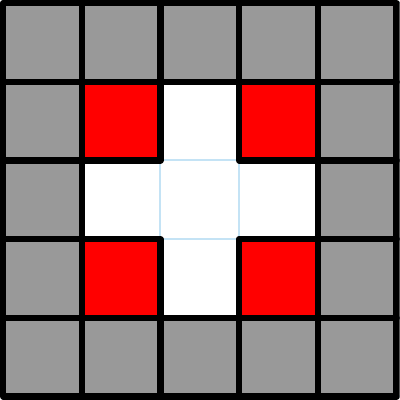}\\
	\caption{2D configuration that cannot be disassembled with a $3$-loose algorithm. The red blocks cannot be removed until at least one other red block is removed.}
	\label{fig:2d-unbuildable}
\end{figure}

Consider the structure in Figure~\ref{fig:2d-unbuildable}. It consists of a hollow $5 \times 5$ square, as well as the four modules in the corners of a concentric $3 \times 3$ square (marked in red). Suppose we can disassemble this structure. Consider the first time we place or remove a module in the central $3 \times 3$ square. We cannot place a module anywhere inside the $3 \times 3$ square, because it would be blocked by red modules. So we must remove a red module. Then this red module must the only module in some $3 \times 3$ square. It is not hard to show that this module cannot have any neighbors, which violates connectivity.

In 3 dimensions, we instead use a hollow $5 \times 5 \times 5$ with $8$ modules in the corners of a $3 \times 3 \times 3$ cube. We can use a similar argument to show that no module can be placed or removed inside the central $3 \times 3 \times 3$ cube.

\section{Universality without Extra Modules}
\label{sec:no-extra}


In this section we show how to reconfigure a modular robot, whose shape satisfies the property of external feature size at least $2$, into a line.
The resulting reconfiguration will consist of 2-accessible and monotone moves (every sliding move will have a sufficient amount of empty space around it, and each module moves only once in a continuous motion from its starting to its target position).
Reversing the sequence moves results in a schedule for the construction of the given structure starting from a line of modules.

The approach we take is similar to the one for the pivoting-cube model, presented in~\cite{M-blocks} and later further developed in~\cite{Feshbach2021,Lo-Wong2021}.
We deconstruct the robot structure slice by slice (parallel to the horizontal plane), ensuring the connectivity of the intermediate structure and the invariant of external feature size at least $2$.
In our algorithm, the choice of which slice to deconstruct is made in the same way as in~\cite{M-blocks,Feshbach2021,Lo-Wong2021}.
The difference of our algorithm with these approaches, and its main difficulty, lies in deciding in which order to deconstruct (partially or fully) a given slice, such that the requirements on the connectivity and the external feature size are satisfied for the intermediate structures.

\begin{figure}
    \centering
    \includegraphics[width=\textwidth]{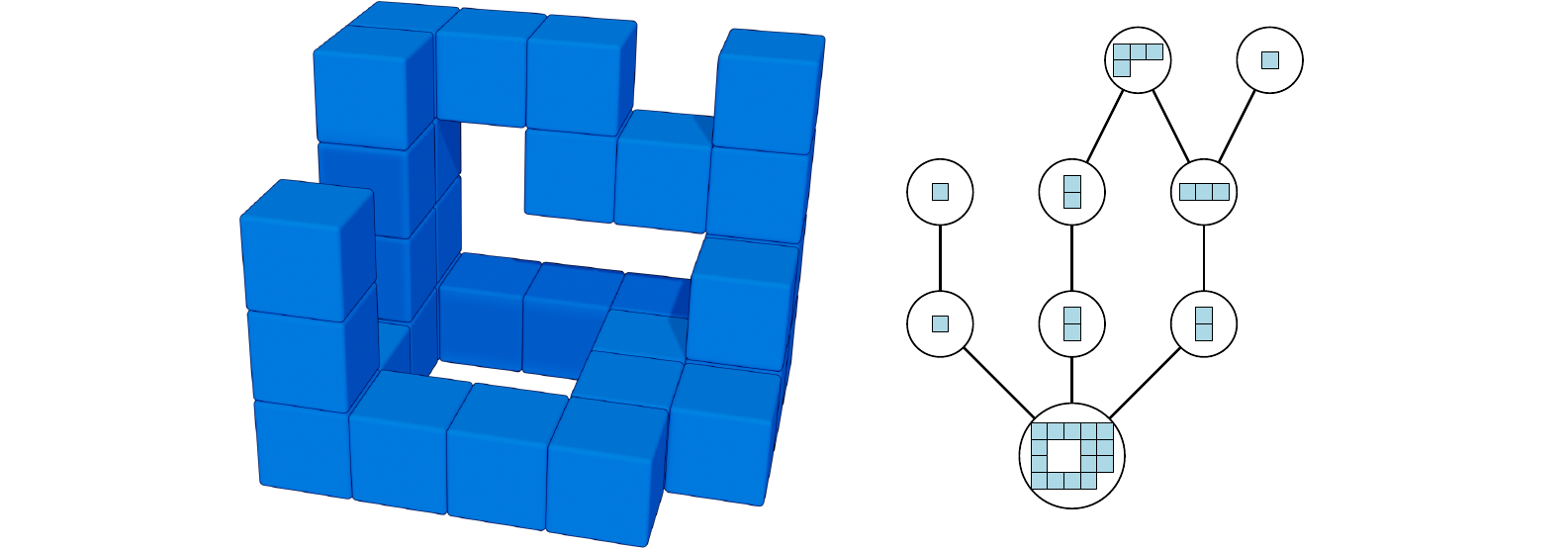}
    \caption{A robot configuration on the left, and the corresponding slice graph on the right.}
    \label{fig:slice-graph}
\end{figure}

\subparagraph{Slice graph.}
We adapt the definition of a slice graph from~\cite{M-blocks}.
Let $V_{z=k}$ denote the subset of nodes of the lattice with the $z$-coordinate fixed to $k$.
Let $V_R$ denote the subset of nodes of the lattice which contain a module in an intermediate configuration $R$.
Define a \defn{slice graph} $G_{s}(V_{s},E_{s})$ of $R$ in the following way.
The nodes $V_{s}$ of $G_{s}$ correspond to the maximally connected components of the modules with the same $z$-coordinate, such that these components have at least one module adjacent to the empty outer space.
That is, each connected component of the induced graphs on $V_{R}\cap V_{z=k}$ (for all $k$), that has a module adjacent to and empty cell of an outer empty space, is added to $V_s$.
Two nodes in $V_s$ are connected with an edge if the corresponding slices contain (vertically) adjacent modules (see Figure~\ref{fig:slice-graph}).

A slice $s\in V_{s}$ is \defn{locally maximal} (\defn{minimal}), if all the adjacent slices of $s$ in $G_{s}$ contain modules with only lower (higher) $z$-coordinate than $s$.

\subparagraph{Outline of the algorithm.}
Consider the initial configuration $T$ with external features of size at least $2$.
Let $s_{0}$ be the slice in $G_{s}$ with the globally maximum $z$-coordinate.
Select an arbitrary module $m_{0}$ from $s_{0}$.
Define a configuration $L$ to be formed by a vertical line of $n$ modules with $m_0$ as the bottom-most module, i.e., all the other modules have the same $x$- and $y$-coordinates as $m_0$, and have $z$-coordinates larger than $z$-coordinate of $m_0$.
We will prove that we can reconfigure from $T$ to $L$ in $O(n^2)$ number of moves.


We iteratively consider a locally extremal outer-surface slice $s$ other than $s_0$ which is not needed for the connectivity of the current slice graph, and move its modules to the top of the line above $m_0$.
There are two cases: either removing $s$ disconnects the structure or it does not. If removal of $s$ does not disconnect the robot structure, we simply completely deconstruct $s$ and proceed with the next locally extremal outer-surface slice.

The harder case is when $s$ disconnects the structure. In this case, $s$ must lie on the boundary of a \defn{void}, an enclosed empty space.
Furthermore, withing this void there must be some modules attached to $s$ which are connected to the rest of the structure only through $s$.
Then, we partially disassemble $s$ while ensuring the connectivity of the intermediate structure, preserving the property of external feature size 2, and connecting the void to the outer empty space.

To summarize, our algorithm consists of the following steps:
\begin{enumerate}
    \item Construct the slice graph $H$;
    \item Using $H$, select a locally extremal outer-surface slice $s$ to deconstruct;
    \item Deconstruct $s$ fully (if connectivity is maintained) or partially, update $H$ accordingly;
    \item Repeat steps 2 and 3 until $L$ has been constructed.
\end{enumerate}

\subparagraph{Step 1: Constructing the slice graph.}
The definition of a slice graph from~\cite{M-blocks} does not require that at least one module from each slice is on the outer surface (adjacent to an empty cell of the outer empty space).
Their result relies on the assumption that any void, a structure to be deconstructed may have, is convex.
Thus, all the slices of the structure have modules that lie on its outer surface.
In particular, all the modules of any locally extremal slice lie on the outer surface.
In our case, however, the structure may have non-convex voids, and thus some slices (including locally extremal ones) may be ``trapped'' inside a void.
We can only move modules that are on the outer surface of the structure.
Thus, to facilitate the choice of a slice to deconstruct, we only maintain the slices in $H$ that are not trapped in a void.
That is, we require the nodes of $G_{s}$ to correspond to the slices with at least one module on the outer surface of the structure.

\subparagraph{Step 2: Selecting a slice to deconstruct.}
We require the slice we select for deconstruction to satisfy two properties: (1)~its removal should not disconnect the remaining structure, and (2)~it should be locally extremal in the $z$-direction to give enough space for the modules to move on its top (or bottom).
Note that naively selecting a slice that corresponds to a non-cut node in $H$ may violate property~(2), and selecting a top-most or bottom-most slice may violate property~(1).

Slice $s_0$, containing the node $m_0$ (the end of the line $L$) will be the last one to be deconstructed.
If there are no other nodes left in $H$, we deconstruct $s_0$.
Otherwise, consider the degrees of the nodes in $V_s$.
If there is a node $s\neq s_0$ with degree $1$, we select it for deconstruction.
Otherwise, select a locally extremal slice $s$ that is furthest away from $s_0$ in $H$.
Removing $s$ from $H$ does not disconnect the graph.
Indeed, otherwise, after the removal of $s$ from $H$, the component not containing $s_0$ would have at least two locally extremal slices.
At least one of these slices existed in $H$ before removing $s$, and is further away from $s_0$ than $s$.
Thus, the lemma follows.
\begin{lemma}\label{slice-connectivity}
If $H$ has more than one slice, there is always a locally extremal slice $s\neq s_0$ in $H$ whose removal does not break the connectivity of $H$.
\end{lemma}

\subparagraph{Step 3: Deconstructing a locally extremal slice.}
Consider a set of empty cells $V_e$ of the empty outer-space component, adjacent to the modules of the structure.
Connect two face-adjacent nodes of $V_e$ with an edge, if the corresponding move is 2-loose or is a convex transition consisting of two 2-loose slide steps.
Observe, that by definition of the external feature size property, graph $G_e$ is connected.
And furthermore, any two nodes of $V_e$ corresponding to face-adjacent cells have an edge connecting them.
\begin{observation}\label{obs:move-to-L}%
Consider a structure $T$ that has external feature size at least $2$, and a top-most node $m_0$.
Consider an additional module $m$ anywhere on the outer surface of $T$.
Then $m$ can move to the top of $m_0$ with loose-sliding moves.
\end{observation}

Let $s$ be our locally extremal slice, and let $z_s$ be the value of the $z$-coordinate of the modules in $s$.
By symmetry, suppose $s$ has no adjacent modules in the positive $z$ direction, thus all the modules adjacent to the modules of $s$ have $z$-coordinate equal to $z_s-1$.

\begin{figure}[b]
    \centering
    \includegraphics[page=2,scale=0.75]{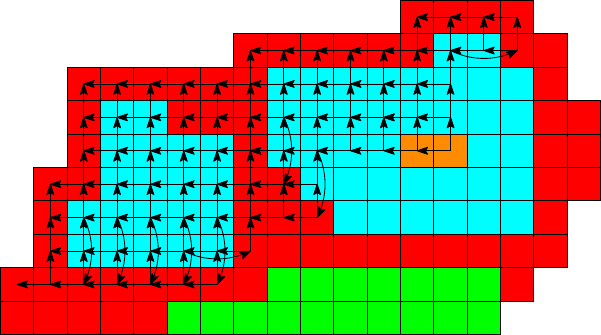}
    \caption{Extremal slice with its modules colored according to the adjacency with the nodes in the slice below. Component of the directed graph $G_s$ of modules reachable from orange modules, used in Lemma~\ref{red-path}.}
    \label{fig:digraph}
\end{figure}

Following the notation of~\cite{Lo-Wong2021}, we assign colors to the modules of $s$, depending on whether they have adjacent modules below or not (refer to Figure~\ref{fig:digraph}). A module $m$ in $s$ is colored:
\begin{itemize}
\item \defn{red} if the cell below $m$ is occupied by some module $m'$, and the slice containing $m'$ is adjacent to the outer empty component,
\item \defn{green} if the cell below $m$ is empty and belongs to the outer component of $R$,
\item \defn{blue} if the cell below $m$ is empty and does not belong to an outer empty component (i.e., it is a void), and
\item \defn{orange} if the cell below $m$ is occupied by some module $m'$, and the slice containing $m'$ is not adjacent to the outer empty component (i.e., it is enclosed in a void).
\end{itemize}


We consider two cases, whether the removal of $s$ from the modular structure breaks its connectivity, or not.
If the removal of $s$ does not break the connectivity of the structure, we move all the modules of $s$ one by one to the end of $L$.
\begin{lemma}
Suppose that removing a locally extremal slice $s$ of $H$ does not disconnect the modular structure.
Then $s$ can be completely deconstructed, and the resulting structure maintains the external-feature-size-$2$ property.
\end{lemma}
\begin{proof}
Consider the graph $G_e$ of empty cells on the surface of the structure, and consider the subset of nodes $V'_e\subset V_e$ that are only adjacent to the modules of $s$ and no other modules.
Observe, that by choice of $s$, removing $V'_e$ from $G_e$ does not disconnect $G_e$.
Thus, to show that we can remove a module from $s$ and move it to the end of $L$, it is sufficient to show how to move it to a slice adjacent to $s$.
From there, there always exists a path to $L$ that does not touch the modules of $s$, and thus does not change through the process of disassembly of $s$.

Let $m_r$ be the right-most top-most red module of $s$, let $x_m$ be its $x$-coordinate, and $y_m$ be its $y$-coordinate.
We can deconstruct the modules of $s$ in the following order.
All the modules with $x$-coordinate larger than $x_m$ must be green.
We deconstruct them column by column, from right to left.
For each of these modules $m_g$, removing $m_g$ maintains the external feature size of at least $2$.
Thus, by Observation~\ref{obs:move-to-L}, $m_g$ can be moved to the end of $L$.

We deconstruct the remaining modules in the decreasing order of their distance (in the dual graph of the modules of the slice) to $m_r$.
Consider a module $m'$ with a maximal distance to $m$.
Clearly, it is in a $2\times 2\times 2$ cube, empty of the modules of the structure other than $m'$, and thus can be removed from its position.
Note that, by the choice of $m'$, the horizontal cycle of empty nodes surrounding $s$ can always be traversed by a module with valid $2$-loose moves.
Thus, we can move $m'$, e.g., clockwise around the slice $s$, until it reaches an empty cell of the outer surface adjacent to $m_r$.
This position is either already adjacent to a slice below $s$, or the module can be moved down once to become adjacent to a slice below $s$.
Similarly to the previous case, from that cell the module can be moved to $L$ along a path that does not touch the modules of $s$.
Thus, all the modules of $s$ can be deconstructed by 2-accessible moves.

Clearly, external feature size 2 and connectivity of the resulting structure after removing $s$ are preserved.
\end{proof}

Consider now the case when removing $s$ disconnects the modular structure.
This is the case if the structure contains a void, and the void has trapped modules inside, attached to $s$.
In this case we no longer can completely deconstruct $s$, and our goal instead is to cut a hole into at least one void without breaking connectivity and while preserving the external feature size of at least $2$.

To do so, we define a directed graph $G_s$ on the modules of $s$.
All pairs of face-adjacent modules in $s$ are connected with a directed edge.
Using $G_s$, we select which modules from $s$ are to be removed at this step of the algorithm.
In particular, all the modules of $s$ that are reachable from orange modules in $G_s$ remain in the slice.
All the other modules are deconstructed.
Below, we argue that this preserves the external-feature-size-2 property, and eliminates at least one void.
Afterwards, we update the slice graph $H$ by adding to it new slices that were trapped inside the eliminated void(s), and proceed to the next iteration of the algorithm.

We construct $G_s$ in the following way.
Add all modules of $s$ as nodes in $G_s$.
We call a module of $s$ \defn{reddish} if it has a module in the layer below it (and so is either red or orange in color), and \defn{bluish} if it does not (and thus is either green or blue).
Add the following edges to $G_s$ (refer to Figure~\ref{fig:digraph}):
\begin{itemize}
\item For two face-adjacent modules with the same $y$-coordinate, add an edge directed to the left.
\item For two face-adjacent modules with the same $x$-coordinate, add an edge directed upward.
\item For a triplet of bluish-bluish-reddish modules oriented left-to-right (or top-to-bottom), add a directed edge from the left-most (top-most) bluish module to the reddish module.
\end{itemize}

\begin{lemma}\label{red-path}
If there are modules present at both locations $a$ and $b$, and $a$ and $b$ are at most 2 units apart in each dimension, then there is a path in the adjacency graph of $s$ from $a$ to $b$ within the bounding box spanned by $a$ and $b$.
\end{lemma}
\begin{proof}
Suppose this is the minimal such counterexample. Consider a sequence of empty cells within this bounding box which starts adjacent to $a$ and ends adjacent to $b$. (If there are no such adjacent empty cells to $a$ and $b$, then this example was not minimal: we could have instead taken one of the occupied cells adjacent to $a$ or $b$ and gotten a smaller counterexample). Now we know that each of these cells is contained in an empty $2\times 2 \times 2$ region. Consider this sequence of $2\times 2\times 2$ regions. It must be possible to slide between each adjacent pair of modules of empty space because we have external feature size 2. However, this is not going to be possible because we cannot fit an empty $2\times 2\times 2$ cube between $a$ and $b$.
\end{proof}

Lemma~\ref{red-path} is the key reason why we use external feature size 2. It follows directly that none of the $1\times 1\times 3$, $1\times 2\times 2$, and $1\times 2\times 3$ forbidden patterns in~\cite{M-blocks} will be present in our structure, and also ensures that every module that is accessible is 2-accessible.
\begin{lemma}
Every minimal path between any module and a blueish module is bitonic in both $x$ and $y$: it never goes right after going left and it never goes down after going up.
\end{lemma}
\begin{proof}
W.l.o.g., consider the shortest such path which starts moving left and ends going right. All of the edges between the first and last edges must be vertical, otherwise there would be a shorter subpath that also satisfies this condition.

If the endpoint of the subpath is above the start, then this path could have been made shorter by removing the first leftward move and the last rightward move, since every cell has a vertically upward edge coming out of it and whenever there is a rightward edge (of length $>1$) which terminates at some cell, there is also a rightward edge starting one unit to the right terminating at the same cell.

Thus our minimal path must have the endpoint below the start. The downward portion of this path must terminate on a reddish cell, because all downward edges do. Call this point $a$. Similarly, the rightward edge at the end of the path must also terminate on a reddish cell. Call this point $b$. Since downward edges never leave reddish cells, there can be at most one downward edge in the vertical portion of the path. This means that $a$ and $b$ must be $\leq 2$ units apart in each dimension. By Lemma~\ref{red-path}, there must be a path of modules in the layer below this one, and thus a path of reddish cells between $a$ and $b$ within their bounding box. In particular, the cell immediately to the right of $a$ is reddish. If $a$ and $b$ are not horizontally adjacent, then it must have been possible to go one space less far left initially, then down to the red cell between $a$ and $b$, and then back up and over to $b$. This is shorter than the original path so our path would not be minimal in this case. If $a$ and $b$ are horizontally adjacent, then consider the next part of the path after $b$ (this is not part of this subpath). There must be such a move because $b$ is reddish and our path ends on a bluish cell. If it goes left before going above the start of our subpath, then we could have shortened the path by skipping the rightward move to $b$. Otherwise it goes up; in this case we had a shorter path by going down first (and not left at all) to the first red below the start of this subpath. Thus in all cases, the path was not minimal, so there cannot be a path with goes left and later goes right.
\end{proof}

\begin{corollary}
All blueish cells reachable from an orange cell are either strictly above it or strictly to the left of it.
\end{corollary}
\begin{proof}
Consider the shortest path to some cell. By construction of $G_s$, the first move must be to the left or up. If it has non-zero leftward component, then it can never go right. If it has non-zero upward component, then it can never go down.
\end{proof}
\begin{corollary}\label{orange-hole}
There is an orange module such that a $2 \times 2$ square directly below and to the right of it is unreachable.
\end{corollary}
\begin{corollary}\label{convex-remove}
If two modules $a$ and $b$ on the same row or column are reachable from an orange module, then every module in between also is.
\end{corollary}

\begin{lemma}\label{2times2edge}
    Any two $2\times 2\times 2$ edge-adjacent empty regions can be slid together until they overlap in at least one cell.
\end{lemma}
\begin{proof}
Suppose for contradiction that we have two such edge-adjacent $2\times 2\times 2$ empty cubes that cannot be slid closer together. This means there is a $2\times 2\times 1$ pattern of diagonally opposite modules with empty space at the other two corners. Suppose we created such a pattern containing one of the modules we removed. Because the region we removed is convex by Corollary~\ref{convex-remove}, it must contain a module from the layer below. In particular, the removed cell from this layer must be red, and there must be an adjacent bluish cell that is not removed. However, in our directed graph there is always an edge from a blue cell to an adjacent red cell, so this is impossible.
\end{proof}

\begin{lemma}\label{blue-square}
    Every removed blue cell is part of a $2 \times 2$ removed square of blue cells.
\end{lemma}
\begin{proof}
    Suppose that a removed blue module is not part of a blue $2 \times 2$ disassembled square. Then it must have two neighbors on opposite sides that are each either reddish or unremoved blue. It cannot have two unremoved blue neighbors on each side because of the convexity of the removed area the algorithm generates. It cannot have two reddish neighbors on each side because that would violate the external feature size beforehand. Finally, it cannot have a unremoved blue pixel on one side and a reddish on the other side because the rules that generate the graph would have connected the blue to the reddish neighbor.
\end{proof}

\begin{lemma}\label{2times2overlap}
    For any pair of overlapping $2 \times 2 \times 2$ empty cubes, at least one of which was created by removing a module from this layer, we can continuously slide one into the other while preserving emptiness. 
\end{lemma}
\begin{proof}
    Because every removed cube is part of a $2 \times 2$ removed square of blue pixels by Lemma~\ref{blue-square}, every new empty space is part of some empty $2 \times 2 \times 2$ cube.

Let the deconstructed layer have $z=z_s$. Consider two overlapping $2 \times 2 \times 2$ empty cubes $\hat Q_1$ and $\hat Q_2$ after partially deconstructing $s$.
If both cubes reside in $z=z_s-1$ and $z=z_s$, then we consider cubes $\hat Q'_1$ and $\hat Q'_2$ that are the projection of $\hat Q_1$ and $\hat Q_2$ to layers $z=z_s-1$ and $z=z_s-2$. Due to the external feature size, $\hat Q'_1$ and $\hat Q'_2$ must both have been empty and it must have been possible to slide between them. We project that motion back up to $\hat Q_1$ and $\hat Q_2$ and get a motion that sweeps between the two cubes.

If both $\hat Q_1$ and $\hat Q_2$ reside in $z=z_s$ and $z=z_s+1$, we consider cases based on the difference in their $x$ and $y$ coordinates.
If they share the same $x$ or $y$ coordinates, then there is a direct motion between them maintaining the shared coordinate.
If one of them is larger in both $x$ and $y$, we know that their bounding box is empty by the convexity of the removed area, so any motion suffices.
If one is smaller in $x$ and larger in $y$, then we slide it first in the increasing $x$ direction, and then in the decreasing $y$ direction.

If, w.l.o.g., $\hat Q_1$ resides in $z=z_s$ and $z=z_s-1$, and $\hat Q_2$ in $z=z_s$ and $z=z_s+1$, then we slide $\hat Q_1$ up one layer first, then proceed with a similar argument to the case above.
\end{proof}

By Lemmas~\ref{2times2edge} and~\ref{2times2overlap}, everything must still have external feature size at least 2.

Finally, by Corollary~\ref{orange-hole}, we have made a hole in slice $s$, which allows us to access modules inside the void.
We have made progress by opening up at least one void. Since at each step of the algorithm we either clear one slice or open at least one void, and there are finitely many voids in the structure, we must eventually terminate with the entire shape deconstructed. Because each module needs to move a distance of at most $O(n)$, and each module is only moved once, the total number of moves is at most $O(n^2)$.

\begin{theorem}
A modular structure with external-feature size at least $2$ can be reconfigured into a line in $O(n^2)$ 2-accessible moves.
\end{theorem}

\section{Open Problems}

The main open question is how many extra modules we need to achieve universal
reconfiguration in the $k$-accessible sliding-cubes model.
Our solution in Section~\ref{sec:extra} uses $\Theta(n)$ extra modules
in the worst case.
But plausibly $O(1)$ extra modules suffice via a different method;
such a result was previously obtained for the pivoting-squares model
\cite{musketeers}.

Another open problem is whether universal reconfiguration is possible for
the $k$-loose sliding-cubes model --- without the $k$-accessible constraint.
In particular, this would require handling complex 3D configurations
with no movable modules on the outside \cite{cubes-multimedia}.
Previous solutions in the sliding-cubes model \cite{BachelorMoreno}
are not loose.

\bibliography{refs}

\end{document}